\documentclass[11pt]{article}
\usepackage{moriond,epsfig}
\usepackage[footnotesize,figbotcap]{subfigure} 

\def\Journal#1#2#3#4{{#1} {\bf #2}, #3 (#4)}

\def\NIMA{{\em Nucl. Instrum. Methods} A}

\def\NPA{{\em Nucl. Phys.} A}
\def\PLB{{\em Phys. Lett.}  B}
\def\PRL{\em Phys. Rev. Lett.}
\def\PRC{{\em Phys. Rev.} C}

\def\CPC{\em Comput. Phys. Commun.}
\def\JPG{{\em J. Phys.} G}

\def\be{\begin{equation}}
\def\ee{\end{equation}}
\def\bea{\begin{eqnarray}}
\def\eea{\end{eqnarray}}

\begin{document}
\vspace*{4cm}
\title{Centrality Dependence of $\Delta\eta - \Delta\phi$ Correlations in Heavy Ion Collisions}

\author{ Edward Wenger, for the PHOBOS collaboration \footnote{The complete PHOBOS collaboration is listed in a recent submission to {\it Physical Review Letters}~\cite{PRL_submission}, from which the text of these proceedings is largely excerpted.} }

\address{Massachusetts Institute of Technology \\
Cambridge, MA, USA}

\maketitle\abstracts{
In these proceedings, a measurement of two-particle correlations with a high transverse momentum trigger particle ($p_T^{trig} > 2.5$~GeV/c) is presented for Au+Au collisions at \mbox{$\sqrt{s_{_{NN}}}$=200 GeV} over the uniquely broad longitudinal acceptance of the PHOBOS detector ($-4 < \Delta\eta < 2$).  As in p+p collisions, the near-side is characterized by a peak of correlated partners at small angle relative to the trigger.  However, in central Au+Au collisions an additional correlation extended in $\Delta\eta$ and known as the `ridge' is found to reach at least $|\Delta\eta| \approx 4$.  The ridge yield is largely independent of $\Delta\eta$ over the measured range, and it decreases towards more peripheral collisions.  For the chosen $p_T^{trig}$ cut of 2.5 GeV/c, the ridge yield is consistent with zero for events with less than roughly 100 participating nucleons.}

\section{Introduction}

Two-particle angular correlations are a powerful tool for elucidating the physics mechanisms responsible for particle production.  Even in environments where the full reconstruction of jets is difficult (e.g. low energy p+p collisions \cite{ISR}), their azimuthally back-to-back signature is manifested in the presence of near-side ($\Delta\phi \approx 0$) and away-side ($\Delta\phi \approx \pi$) peaks in the azimuthal correlation of particles associated with the leading hadron.  

In heavy ion collisions, correlation measurements with respect to the leading hadron have provided a great deal of insight into the properties of the produced medium.  In particular, the disappearance of back-to-back high $p_T$ correlated pairs \cite{STAR_BackToBackJets} is interpreted as evidence for an opaque medium that induces large energy losses for partons traversing the interior.  

By lowering the $p_T$ threshold for associated particles in the correlation function, one can study where the energy is lost and how the medium responds to penetrating high $p_T$ probes.  In such studies at mid-rapidity, the STAR collaboration \cite{STAR_Ridge} has observed an enhancement of two-particle angular correlations near $\Delta\phi\approx0$ (called the `ridge') extending at least to \mbox{$|\Delta\eta|<1.7$}.  To distinguish among the many possible theoretical interpretations of this phenomenon, the PHOBOS experiment \cite{PRL_submission} has measured the correlated yield of charged hadrons with respect to a high transverse momentum trigger particle (\mbox{$p_T > 2.5$}~GeV/c) up to large relative pseudorapidity ($|\Delta\eta|\approx4$).

\section{Analysis}

The PHOBOS detector \cite{PHOBOS_NIM} has two azimuthally opposed spectrometer arms, each covering 0.2 radians in $\phi$, which are used to select charged trigger tracks with \mbox{$p_T>2.5$}~GeV/c within an acceptance of \mbox{$0<\eta^{trig}<1.5$}.  Associated particles that escape the beam pipe ($p_T^{min} \simeq 35 \rightarrow 7$~MeV/c for $\eta=0 \rightarrow 3$) are detected in a single layer of silicon comprising the octagon subdetector ($|\eta|<3$). 

The construction of the per-trigger correlated yield of charged particles is described in any given centrality class by the following expression:

\begin{equation}
 \frac{1}{N_{trig}} \frac{d^{2}N_{ch}}{(d\Delta\phi) (d\Delta\eta)} = B(\Delta\eta)  \cdot \left[\frac{s(\Delta\phi,\Delta\eta)}{b(\Delta\phi,\Delta\eta)}  - a(\Delta\eta) [ 1+2V(\Delta\eta) cos(2\Delta\phi) ]\right].
\label{eqn:CorrelatedYield}
\end{equation}

\noindent The raw per-trigger distribution of same-event pairs $s(\Delta\phi,\Delta\eta)$ is divided by the raw mixed-event background distribution $b(\Delta\phi,\Delta\eta)$ to account for random coincidences and acceptance effects.  This ratio is calculated in $1$~mm bins of vertex position along the beam and averaged over the range $-15<z<10$~cm.  
$B(\Delta\eta)$, which converts the flow-subtracted correlation into a conditional yield, is constructed from the single-particle distribution ($dN/d\eta$) \cite{dNdeta} by convolution with the normalized $\eta$ distribution of trigger particles. 

Because elliptic flow is erased in the mixing of tracks and hits from different events, the remaining $\langle v_2^{trig} v_2^{assoc}\rangle$ modulation carried by $s(\Delta\phi,\Delta\eta)$ must also be subtracted.  The flow magnitudes are calculated from a parameterization based on published PHOBOS measurements of $v_2$ as a function of  $N_{part}$, $p_T$, and $\eta$ \cite{PHOBOS_Flow}, assuming a factorized form.    

\begin{figure}[htbp]
\centering
	   \includegraphics[width=0.65\textwidth, viewport=20 10 550 300]{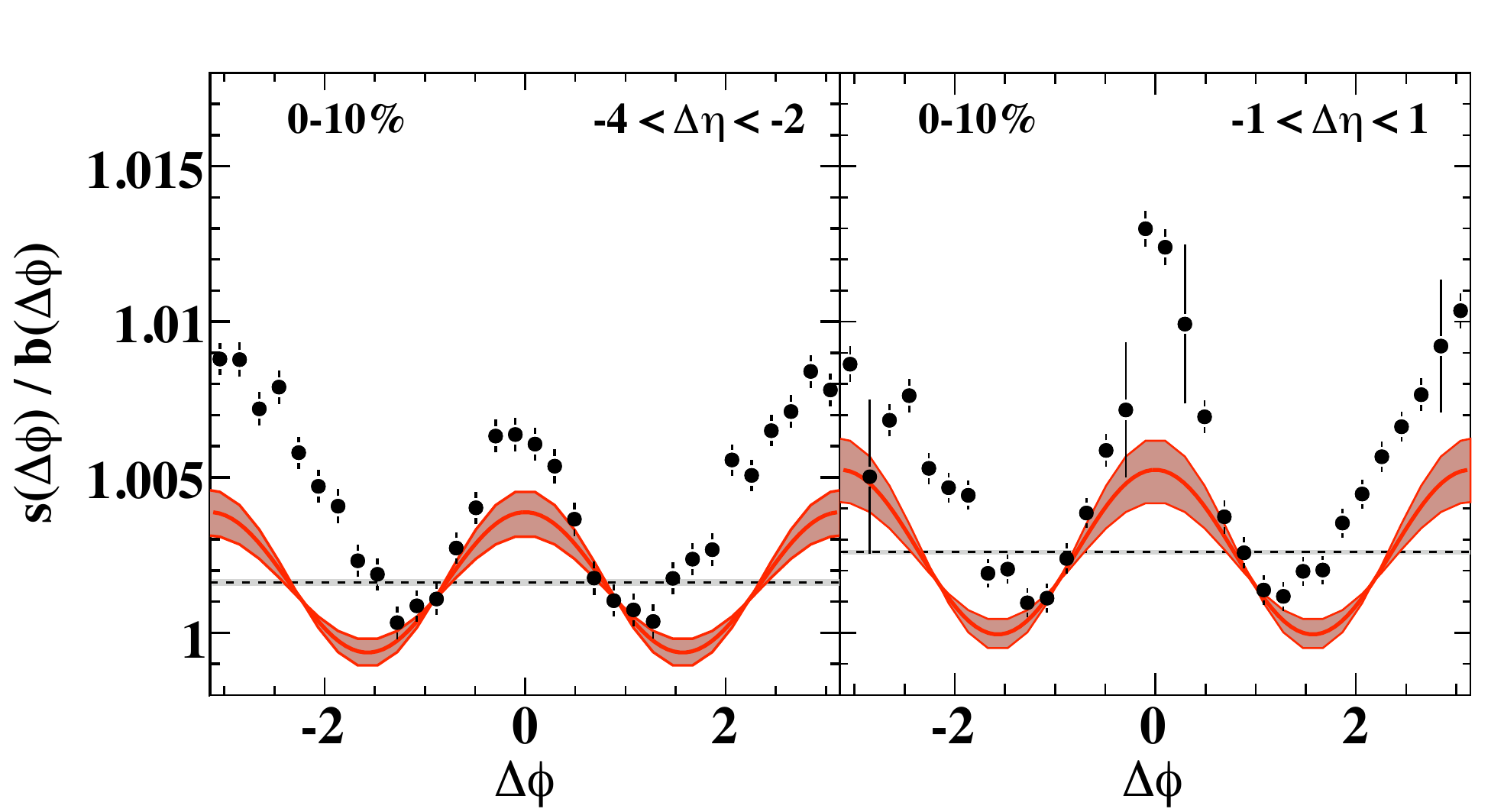}
\caption{Ratio of signal to background for the 10\% most central collisions at long-range (left, $-4<\Delta\eta<-2$) and short-range (right, $|\Delta\eta|<1$).  The estimated flow modulation (scaled by $a(\Delta\eta)$) and its uncertainty are represented by the solid line and shaded band.  The uncertainty on the ZYAM parameter is represented by the narrow band around the dashed line.}
\label{fig:flowsubtraction}
\end{figure}

The scale factor $a(\Delta\eta)$ is introduced to account for the small difference in multiplicity between signal and mixed-event distributions.  Its value -- within a few percent of unity in all cases considered -- is extracted separately in bins of centrality and $\Delta\eta$, using the modified implementation of the zero yield at minimum (ZYAM) method described in Alver {\it et al} \cite{PRL_submission}.
The subtraction of elliptic flow from the raw correlation is illustrated in Fig.~\ref{fig:flowsubtraction}. 

\section{Results}

\begin{figure}[htbp]
\centering
   \subfigure[~p+p PYTHIA]{
   	\includegraphics*[width=.48\textwidth, viewport=20 10 550 360]{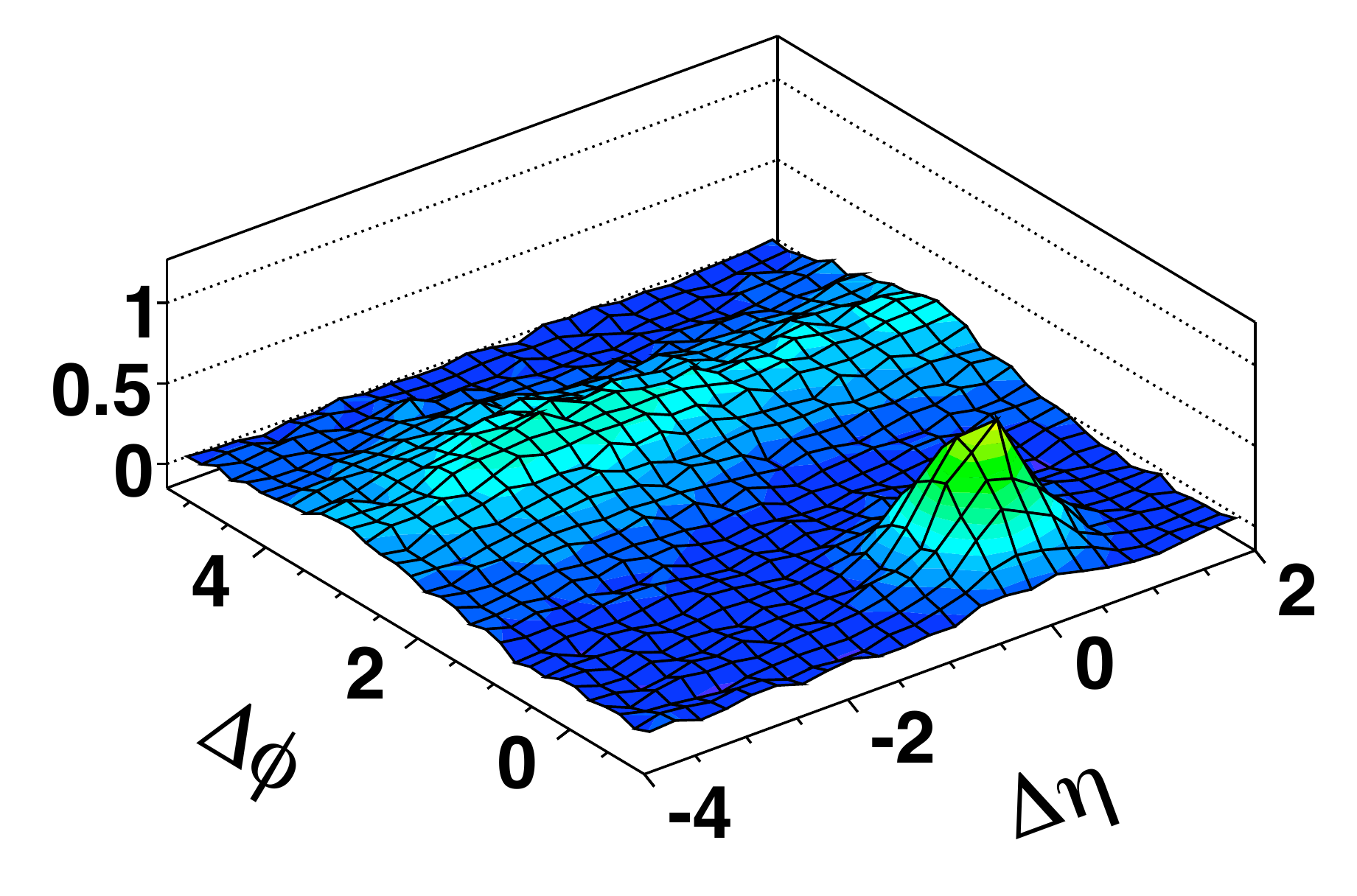} 
	\label{fig:pythiacorr}
   }
   \subfigure[~Au+Au 0-30\% (PHOBOS)]{
   	\includegraphics*[width=.48\textwidth, viewport=20 10 550 360]{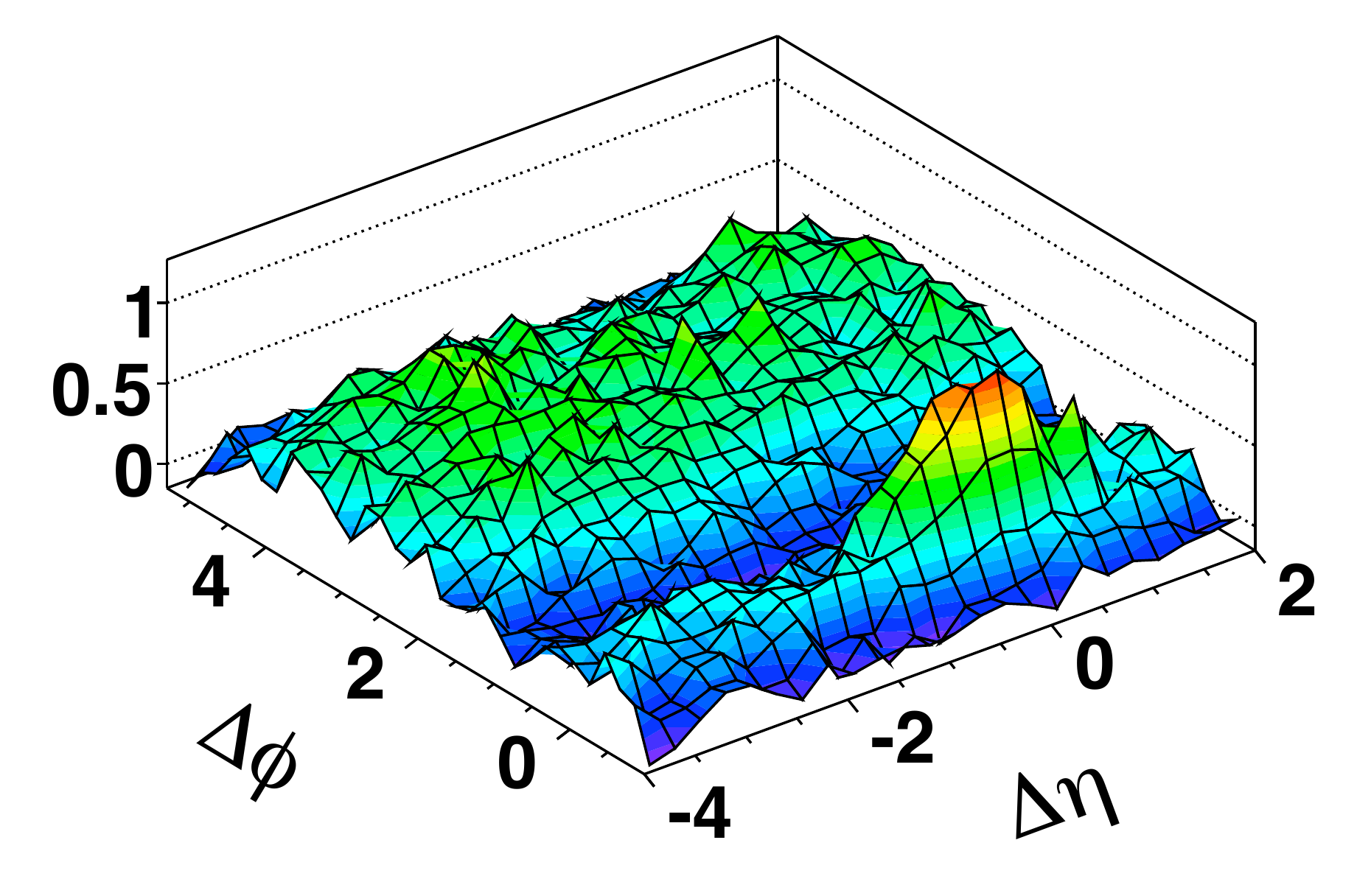} 
	\label{fig:auaucorr}
   }

\caption {
Per-trigger correlated yield with \mbox{$p_T^{trig} > 2.5$}~GeV/c as a function of $\Delta\eta$ and $\Delta\phi$ for  
 $\sqrt{s}$ and $\sqrt{s_{_{NN}}}$=200 GeV \subref{fig:pythiacorr} PYTHIA p+p and \subref{fig:auaucorr} PHOBOS \mbox{0-30\%} central Au+Au collisions.
}
\label{fig:corrsurf}
\end{figure}

To understand the effects of the hot, dense medium on correlated particle production, the PHOBOS Au+Au data is compared to p+p events simulated with PYTHIA version 6.325 \cite{Pythia}.  The prominent features of the p+p correlation, shown in Fig.~\ref{fig:pythiacorr}, are a jet-fragmentation peak centered at $\Delta\phi=\Delta\eta=0$ and an away-side structure centered at $\Delta\phi = \pi$ that is similarly narrow in $\Delta\phi$ but extended in $\Delta\eta$, since the hard scattering can involve partons with very different fractions of the proton momentum.  

In central Au+Au collisions, particle production correlated with a high $p_T$ trigger is strongly modified as shown in Fig.~\ref{fig:auaucorr}.  Not only is the enhanced away-side yield much broader in $\Delta\phi$, the near-side peak at \mbox{$\Delta\phi\approx0$} now sits atop a pronounced ridge of correlated partners extending continuously and undiminished all the way to $\Delta\eta=4$.  

\begin{figure}[htbp]
\centering
\subfigure[~Near-side $\Delta\eta$ projection ($|\Delta\phi| < 1$)]{
	\includegraphics*[width=.48\textwidth]{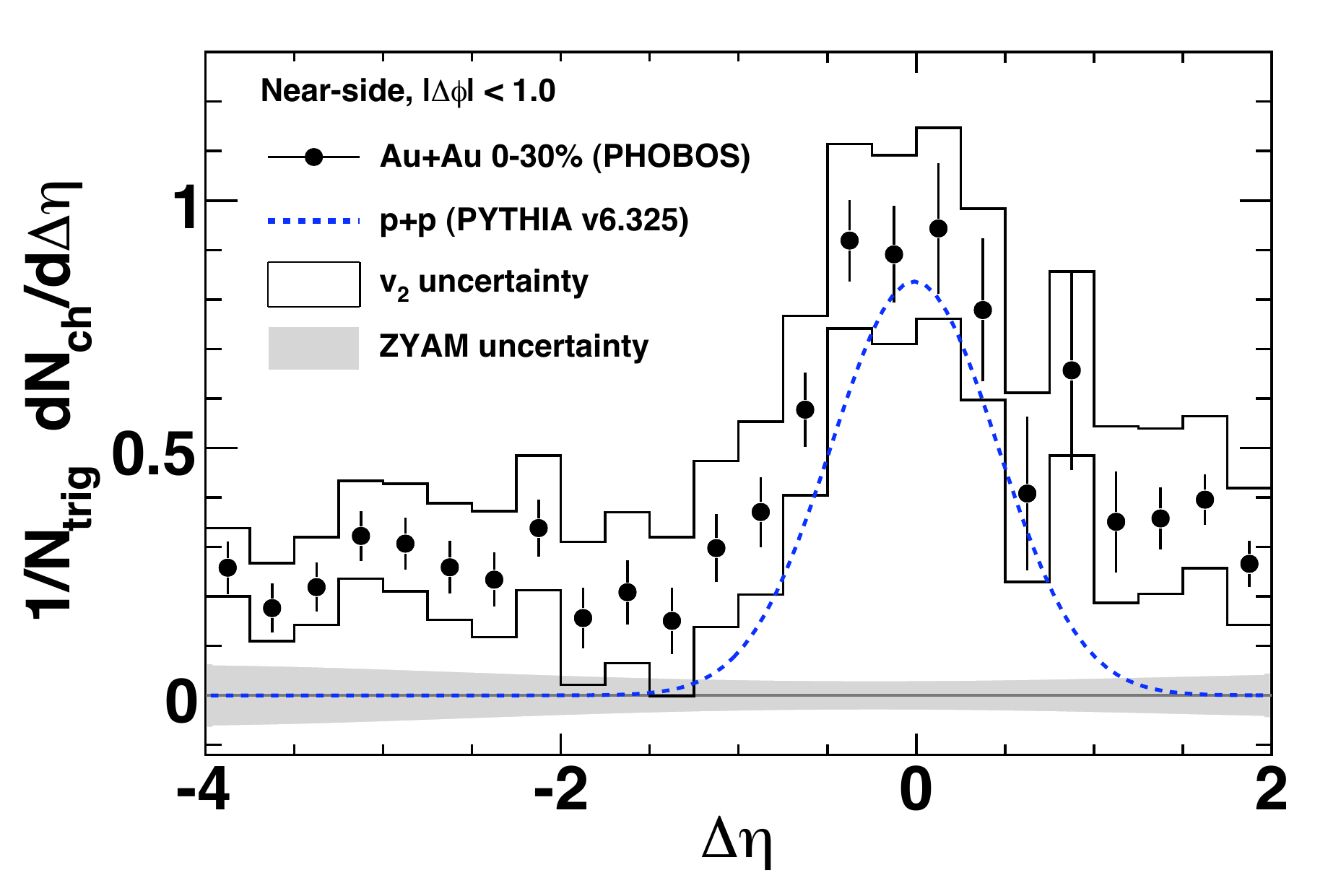}
	\label{fig:detaproj}	
}
\subfigure[Integrated yields versus centrality]{
	\includegraphics[width=0.48\textwidth, viewport=20 20 550 420]{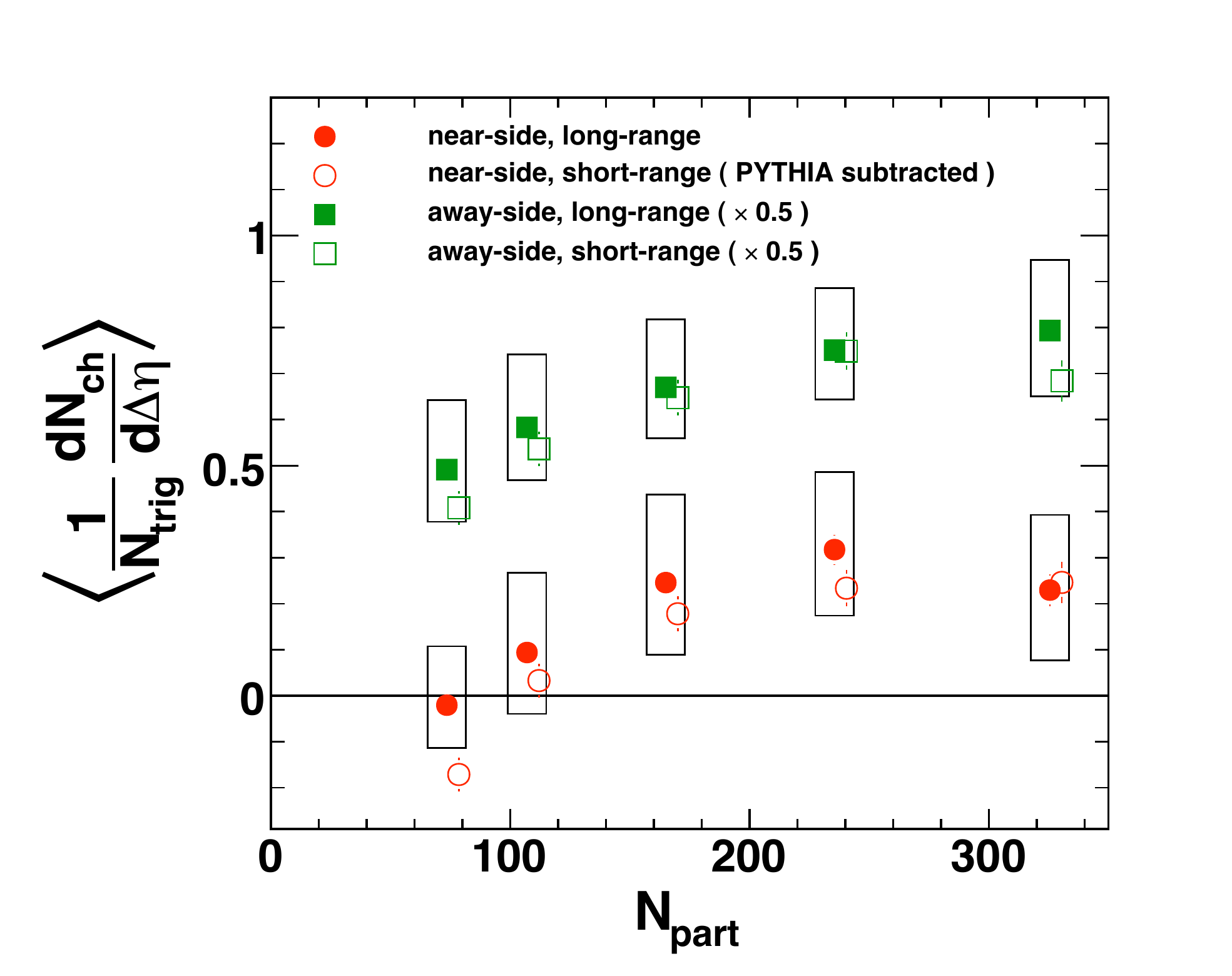} 
	\label{fig:RidgeVsNpart}
}
\caption{\subref{fig:detaproj} Near-side yield integrated over \mbox{$|\Delta\phi| < 1$} for 0-30\% Au+Au compared to PYTHIA p+p (dashed line) as a function of $\Delta\eta$.  Black boxes represent the uncertainty from flow subtraction. The error on the ZYAM procedure is shown as a gray band at zero. All systematic uncertainties are 90\% confidence level. 
\subref{fig:RidgeVsNpart} Average near-side ($|\Delta\phi|<1$) and away-side ($\Delta\phi>1$) yields as a function of $N_{part}$ at both short-range ($| \Delta\eta| < 1$) and long-range ($-4 < \Delta\eta < -2$).  The near-side yield at short-range has the PYTHIA jet yield subtracted from it.  For clarity of presentation, the open points have been offset slightly and the away-side yields scaled down by a factor of 2.  Boxes correspond to the combined 90\% systematic uncertainties on the $v_2$ estimate and ZYAM procedure.
}
\end{figure}

To examine the near-side structure more closely, the correlated yield is integrated over the region $|\Delta\phi|<1$ and plotted as a function of $\Delta\eta$ in Fig.~\ref{fig:detaproj}.  For the most central 30\% of Au+Au collisions, there is a significant and relatively flat correlated yield of about 0.25 particles per unit pseudorapidity far from the trigger.  

The similarity of the short-range ($| \Delta\eta| < 1$) and long-range ($-4 < \Delta\eta < -2$) yields in excess of the p+p jet correlation suggests a decomposition of the near-side correlation into distinct jet and ridge components.  Such a separation is supported by previous STAR measurements of the associated particle $p_T$ spectra, the centrality independence of the jet-like yield, and the ridge-subtracted fragmentation function \cite{STAR_Ridge}.

In Fig.~\ref{fig:RidgeVsNpart}, the integrated ridge yield at long-range (filled circles) is compared to the short-range yield (open circles) after subtraction of a jet component corresponding to the PYTHIA yield.
Already for this simple assumption, the ridge yield is shown to be the same within experimental uncertainties at all $\Delta\eta$.  It decreases as one goes towards more peripheral collisions and is consistent with zero in the most peripheral bin analyzed (40-50\%).  While the systematic errors do not completely exclude a smooth disappearance of the ridge as one approaches p+p collisions, these data suggest that with a $p_T = 2.5$~GeV/c trigger particle the ridge may already have disappeared by $N_{part}=80$ (about 45\% central).

\section{Summary}

In these proceedings, the PHOBOS measurement of the ridge at small $\Delta\phi$ has been presented over a broad range of $\Delta\eta$.  The fact that in central collisions the ridge extends to at least four units of pseudorapidity away from the trigger is a powerful constraint on theories that strive to explain the nature of particle production in heavy ion collisions.  In particular, causality considerations require that the correlation be imprinted in the very earliest moments after the collision \cite{Dumitru}.  Taken together with the other known properties of the ridge -- its makeup is similar to the bulk but not jet fragmentation \cite{STAR_Ridge}; it is observed in inclusive correlations without a high $p_T$ trigger \cite{Daugherity} -- the longitudinal extent of ridge correlation suggests an interpretation other than medium-induced broadening of jet fragmentation \cite{Armesto}.  
More quantitative theoretical studies will be required to determine which proposed mechanisms can consistently describe the broad extent of the ridge and its dependence on collision geometry.

\section*{Acknowledgments}
This work was partially supported by U.S. DOE grants 
DE-AC02-98CH10886,
DE-FG02-93ER40802, 
DE-FG02-94ER40818,  
DE-FG02-94ER40865, 
DE-FG02-99ER41099, and
DE-AC02-06CH11357, by U.S. 
NSF grants 9603486, 
0072204,            
and 0245011,        
by Polish MNiSW grant N N202 282234 (2008-2010),
by NSC of Taiwan Contract NSC 89-2112-M-008-024, and
by Hungarian OTKA grant (F 049823).


\section*{References}
\begin{thebibliography}{99}

\bibitem{PRL_submission}B. Alver {\it et al}, arXiv:0903.2811 (2009).

\bibitem{ISR} A.L.S. Angelis {\it et al}, \Journal{\PLB}{97}{163}{1980}.

\bibitem{STAR_BackToBackJets} C. Adler {\it et al}, \Journal{\PRL}{90}{082302}{2003}.

\bibitem{STAR_Ridge} J. Putschke, \Journal{\JPG}{34}{S679}{2007}.

\bibitem{PHOBOS_NIM}B.B. Back {\it et al}, \Journal{\NIMA}{499}{603}{2003}.

\bibitem{dNdeta}B.B. Back {\it et al}, \Journal{\PRL}{91}{052303}{2003}.

\bibitem{PHOBOS_Flow}B.B. Back {\it et al}, \Journal{\PRC}{72}{051901}{2005}.

\bibitem{Pythia}T. Sj\"ostrand {\it et al}, \Journal{\CPC}{135}{238}{2001}.

\bibitem{Dumitru}A. Dumitru {\it et al}, \Journal{\NPA}{810}{91}{2008}.

\bibitem{Daugherity}M. Daugherity, \Journal{\JPG}{35}{104090}{2008}.

\bibitem{Armesto}N. Armesto {\it et al}, \Journal{\PRL}{93}{242301}{2004}.

\end{thebibliography}
\end{document}